# Identification and Analysis of Transition and Metastable Markov States

Linda Martini,[1] Adam Kells,[1] Gerhard Hummer,[2] Nicolae-Viorel Buchete,[3] and Edina Rosta[1]

[1]*King's College London, Department of Chemistry, SE1 1DB, London, UK*

e-mail address: edina.rosta@kcl.ac.uk

[2]*Max Planck Institute for Biophysics, Department of Theoretical Biophysics, Frankfurt, 60438, Germany*

[3]*University College Dublin, School of Physics, Dublin 4, Ireland*

We present a new method that enables the identification and analysis of both transition and metastable conformational states from atomistic or coarse-grained molecular dynamics (MD) trajectories. Our algorithm is presented and studied by using both analytical and actual examples from MD simulations of the helix-forming peptide $Ala_5$, and of a larger system, the epidermal growth factor receptor (EGFR) protein. In all cases, our method identifies automatically the corresponding transition states and metastable conformations in an optimal way, with the input of a set of relevant coordinates, by capturing accurately the intrinsic slowest relaxation rate. Our approach provides a general and easy to implement analysis method that provides unique insight into the molecular mechanism and the rare but crucial rate limiting conformational pathways occurring in complex dynamical systems such as molecular trajectories.

# I. INTRODUCTION

Recent advances in both parallelizable computational software and the development of highly-efficient supercomputers have extended the timescale accessible to atomistic molecular dynamics (MD) with explicit solvent representations to simulations up to the order of milliseconds (1-3). While this enables state-of-the-art computational modeling of complex molecular processes such as folding and binding (4), the vast amount of complex, high-dimensional data obtained from these simulations requires novel analysis methods, on one hand, to make use of all the available information and, on the other hand, to extract comprehensible and relevant information. The use of Markov State Models (MSMs, (5, 6)) allows to directly extract thermodynamic and kinetic information from MD trajectories, such as free energy profiles, equilibrium probabilities and transition rates, and has proven to be a useful tool to analyze data from both experiments and simulations (7-11).

A variety of methods have been proposed that attempt to aggregate the conformational state space, projected along certain reaction coordinates (RCs) of interest into macrostates, proposing approaches that can be implemented automatically (12-14). Most of these methods rely on kinetic network descriptions employed by MSMs. However, while clustering algorithms such as PCCA+ (15) and its advances (16, 17) are designed to identify metastable states (e.g., in protein folding terms: the folded, the unfolded and long-lived intermediate states), none is aimed specifically at the automatic, reliable identification and characterization of the rate-determining transition states (TS), which are crucial for the thorough understanding of underlying molecular mechanisms and which control the overall kinetics of the system. A method that allows the identification of TS, defining them as unstable states with equal probability to transition to neighboring metastable states on either side along the reaction coordinate, while concurring maximum flux, makes use of the commitment probability of states (18) (i.e. in one dimension, the probability to reach one state before it reaches another). However, in complex processes, it is not clear how many metastable states are important to describe the overall slow dynamics, and where the key kinetic bottleneck is located.

Here we propose a method for the automatic identification and analysis of both metastable conformational states (MS) and TS regions, by optimal and hierarchical construction of MSMs using a set of discretized RCs. The use of key RCs has long been a successful approach in many of the major enhanced sampling methods (19-22), and it provides a crucial framework to allow the intuitive understanding of the data (23). Our algorithm enumerates all possible clusters that could be formed along an ordered set of states for each RC, and selects the optimal clustering that maximizes the slowest relaxation time of the reduced system. This is both a physically and mathematically meaningful optimization process, as the full system's relaxation time provides an upper bound to the slow relaxation processes of this system, and the better the coarse graining, the closer our objective function will be to this ideal value. Furthermore, our algorithm does not require higher order eigenvectors besides the second eigenvalue (unlike e.g., PCCA+). By increasing the number of coarse-grained states used to describe the system, we systematically identify key MSs, until we exhaust the number of relevant MSs, and the first optimal TS is identified. This approach helps define the minimal required number of MSs relevant for the kinetics of the process. We subsequently combine all the RCs in corresponding direct product states to finally obtain a reduced kinetic model of the system described by the complete set of RCs, and the corresponding TS that is most relevant to the slowest intrinsic relaxation time (24).

## II. TIMESCALE OPTIMIZATION CLUSTERING

Based on the fact that any clustering of MSs decreases the measured relaxation times of the intrinsic dynamics of a system (25), and considering our main aim of estimating most accurately the slowest relaxation process, we target our proposed clustering method towards the maximization of the slowest extracted relaxation time (15, 24).

### A. 1D Clustering

Complex, multidimensional trajectories are often analyzed using their projection on simpler 1D RCs. Here we start by assuming that a set of degrees of freedom is available, which accounts for the slow dynamics of the system. Any continuous 1D RC can be discretized into adequately small "*microstates*" (μ-states), denoted here as $s_1, ..., s_{N_\mu}$, where $N_\mu$ is their total number, which capture the same intrinsic dynamics as the continuous RC. In practice, a 1D RC, $x(t)$, is often binned in equidistant μ-state intervals over RC windows of length $\Delta x_\mu$, such that, for a trajectory that samples the interval $[x_{min}, x_{max}]$ we have $N_\mu = (x_{max} - x_{min})/\Delta x_\mu$. This discretization has an intrinsic ordering inherited from the continuous RC.

To reduce the dimensionality, the typically large number of $N_\mu$ μ-states can be clustered into coarse-grained "*macrostates*" (M-states). Given the projection of a trajectory on a 1D RC, $x(t)$, for clustering it into $M$ sorted M-states over a finite domain (*i.e.*, with no periodic boundaries), $b_i$ cluster boundaries ($i \in \{1, 2, ..., M-1\}$) are needed. Discretizing $x(t)$ into $M$ macrostates in the interval $[x_{min}, x_{max}]$ requires $M - 1$ boundaries. Any positioning of the boundaries defines a distinct set of M-states along the RC: $S_0 = [x_{min}, b_1), S_1 = [b_1, b_2), ..., S_{M-2} = [b_{M-2}, b_{M-1}), S_{M-1} = [b_{M-1}, x_{max}]$, and we are searching for the one that maximizes the coarse grained relaxation time ( $\max\{t_2 \mid b_1,\ldots,b_{M-1}\}$ ).

When the exact full rate matrix **K** is known (i.e., not only the transition probabilities from trajectories), the reduced rates for each choice of a set of state boundaries, $b_i$, along a Markov chain can be obtained using optimal, lagtime-independent expressions (e.g., Eq. 12 in Ref. (24)). When the full rate matrix **K** is not available, approximate reduced Markov matrices can be built, see for example Fačkovec et al. (10). Here we considered a lag time τ, for which the corresponding transition count matrix **C**(τ) can be calculated, with elements $C_{ij}(\tau)$ corresponding to the number of transitions observed from $S_i$ to $S_j$ during the observation window τ. Subsequently, the corresponding Markovian transition probability matrix can be built, with entries $T_{ij}(\tau) = \dfrac{C_{ij}(\tau)}{\sum_{j=1}^{M} C_{ij}(\tau)}$ or using a reversible estimator with an iterative approach (26-29). Solving the eigenvalue problem of the Markov matrices **T**(τ) for each possible set of boundaries, the slowest relaxation time is obtained using $t_2 = -\tau/\log \lambda_2$, where $\lambda_2$ is the second largest eigenvalue, and $\lambda_1 = 1$. Finally, the optimal *M*-state clustering corresponds to the *M*−1 boundary positions, for which the slowest relaxation time, $t_2$, is maximum.

Importantly, by increasing the number of M-states one-by-one, we first obtain all the key MSs, corresponding to diagonally-dominated reduced transition probability matrices ($T_{ij}^\tau$). Our algorithm automatically evaluates the transition probabilities to neighboring states and these are compared with the probability to remain in the same state. A TS is defined as an unstable state from which the outgoing probabilities to either directions are greater than the probability to stay in the same state, thus in our implementation, as we identify new M states, we test for this property and we stop adding additional M states once a TS is found. In addition, we also verify the nature of our TS by calculating and ensuring that most of the reactant flux goes through these states (see SI Section I) rather than around them, in accordance with the maximum flux criteria proposed by Huo and Straub (30). This characteristic qualitative change in the properties of the last M-state allows us to identify the minimal number of key MSs that have to be accounted for in the description of the process along each 1D RC, and more generally in *N*-D.

### B. *N*-D Clustering

To generalize our approach to the analysis of a trajectory in the *N*-D state space (i.e., described by *N* 1D RCs), each dimension is first clustered into *M* macrostates as described above. Assuming for simplicity that the number of M-states is the same, *M*, for each dimension (an assumption that is relaxed later, with no loss of generality), a 1D trajectory with $M^N$ states is obtained. We can thus define $M^N$ unique states (e.g., of the form $\Sigma_0 = [00000] = \{S_0, S_0, S_0, S_0, S_0\}$ when the trajectory is in the M-state $S_0$ in all $N = 5$ dimensions). The resulting $M^N$ macrostates, are relabeled from M-state $\Sigma_0$ to M-state $\Sigma_{M^N-1}$, and define our 1$^{st}$-level coarse-grained trajectory.

Finally, to perform a 2$^{nd}$-level coarse graining, the M-states $\Sigma_0, \ldots, \Sigma_{M^N-1}$ are first ordered (e.g., by their commitment probability values, calculated from the right eigenvector corresponding to the $\lambda_1 = 1$ eigenvalue of the Markov model with $M^N$ states (15, 18, 24)) for a computationally feasible algorithm analogously to the one presented by Hummer and Szabo (24). We note here that the commitment probability cannot in general be used to define transition states in complex systems with multiple states (see also Figure S2). Instead, we use the second eigenvector here only to order the states, and we define subsequently the boundaries between the coarse-grained states based on this ordering initially. The resulting, sorted M-state trajectory are then analyzed one more time according to the 1D clustering procedure described above, to find a smaller and coarser grained optimal clustering into $L$ larger "$\Omega$-states" (i.e., with a maximum corresponding slowest relaxation time, $1/\lambda_2$). The resulting 2$^{nd}$-level coarse graining into $\Omega$-states (i.e., with $2 < L < M^N$) is denoted here by $\Omega_1, \ldots, \Omega_L$. The ordering according to the second right eigenvector is not optimal in all cases, however, we verified in all examples presented here, that single state variations do not significantly affect the obtained clustering results. We also note that by using an ordering along the second right eigenvector, the continuity of the states is no longer ensured, and therefore kinetically disconnected states might be directly adjacent to one another. This problem may be overcome by using alternative approaches, for example diffusion maps (31) to identify an ordering that is based on kinetic vicinity of the states.

This final coarse-graining procedure is presented above by considering all the RCs at the same time, globally. However, importantly, for complex systems with a large number of RCs it may be necessary to perform the final clustering step by hierarchically approximating the coarse graining of the M-states, into the $\Omega$-states, $\Omega_1, \ldots, \Omega_L$, by clustering the M-states, $S_0, \ldots, S_{M-1}$, of only a few (e.g., two) RCs at a time in a stepwise manner, and include additional RCs sequentially. This avoids the clustering of an unfeasibly large number of M-states $\Sigma_0, \ldots, \Sigma_{M^N-1}$ at the same time.

### C. Analytical Investigation

We first tested our clustering method using Markov state models created for a set of analytical free energy functions that can be tuned to switch continuously and monotonically between 2-state-like and

3-state-like dynamics. To define the kinetic model underlying the analytical free energy function $F(x)$, we construct an $N_\mu$-state Markov chain. The corresponding rate matrix, **K**, is given by:

$$(1a) \quad K_{i,i+1} = A \exp\left[\frac{F(x_i) - F(x_{i+1})}{2k_B T}\right]$$

$$(1b) \quad K_{i,i} = -\sum_{\substack{j=1 \\ j \neq i}}^{N_\mu} K_{i,j}.$$

$A$ is the Arrhenius prefactor (here, $A=10$), $k_B$ is Boltzmann's constant, and $T$ is the absolute temperature (here $k_B T = 0.596$).

Our general aim is to obtain a clustering of the underlying Markov chain into $M$-state aggregates $S_0,\ldots,S_{M-1}$ defined by optimal cluster boundaries $b_1,\ldots,b_{M-1}$ that correspond to a maximum slowest relaxation time.

For example, a simplest 3-state coarse graining leads to a reduced 3x3 rate matrix

$$(2) \quad K = \begin{bmatrix} -K_{12} & K_{21} & 0 \\ K_{12} & -(K_{21}+K_{23}) & K_{32} \\ 0 & K_{23} & -K_{32} \end{bmatrix},$$

corresponding to the theoretical case of a linear chain of states (e.g., Fig. 1.), with elements depending on the coarse graining boundaries. Optimal boundaries are obtained by minimizing the magnitude of the resulting $\lambda_2$ eigenvalue of **K** (SI, Section II.), which generally corresponds to a separation into MSs. However, for 2-state-like systems, the solution of the eigenvalue optimization problem will lead to boundaries that identify a TS. Thus, when the 2$^{nd}$ state has a TS-like character, this corresponds to reduced rates that approximate $\lambda_2$ (see also SI, Section II.) as

$$(3) \quad -\lambda_2 \approx \frac{K_{32}K_{21} + K_{12}K_{23}}{K_{21} + K_{23}}.$$

Here, we calculated the reduced Markov matrix both using Eq. 12 of Hummer and Szabo (24), and also with respect to a lag time $\tau$ using (see also SI Section III.):

$$(4) \quad T_{ij}^\tau = \left(\pi_i^{eq} \exp(\mathbf{K}\tau)\right)\mathbf{e}_j = \langle \mathbf{e}_j | \exp(\mathbf{K}\tau) | \pi_i^{eq} \rangle$$

With $\mathbf{e}_i = (0,\ldots,0,1,\ldots,1,0,\ldots,0)^T$, with non-zero entries from indices $b_{i-1}+1$ to $b_i$, $\pi_i^{eq} = (0,\ldots,0,p_{b_{i-1}+1}^{eq},\ldots,p_{b_i}^{eq},0,\ldots,0)/\left(\sum_{k=b_{i-1}+1}^{b_i} p_k^{eq}\right)$ where $\mathbf{p}^{eq}$ is the left eigenvector of **K** corresponding to eigenvalue 0.

Varying the cluster boundaries $b_i$ and maximizing the slowest relaxation time for all reduced Markov matrices defines the clustering algorithm for the analytical model, with the parameters $\tau$ and $N_\mu$.

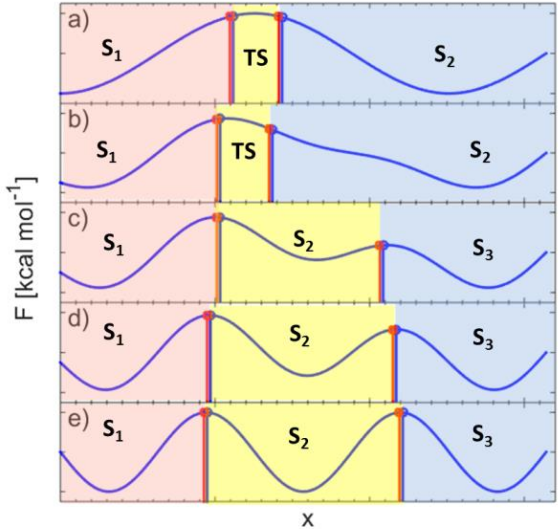

**Figure 1.** Illustration of automatic optimal clustering into $M=3$ macrostates for a set of 1D analytical potentials (panels **a** to **e**, see SI Section III.) in which the intrinsic kinetics is tuned continuously as a function of a control parameter from being 2-state-like (**a**) to 3-state-like (**e**). The middle region (yellow) between the two boundaries identified by the algorithm (vertical lines) is either a TS, in **a** and **b**, or it becomes the third MS in **c-e**. Vertical lines correspond to the Hummer and Szabo rates (blue, (24)), and to transition probability matrices at $\tau = 1000$ (red).

Application of this method to a multitude of free energy profiles (Fig. 1) revealed that the slowest timescale optimization clustering successfully identifies all meta-stable states in a system. Once we exhausted the number of available important MSs as explained in Section IIIA, the next optimal M-state is qualitatively different from the metastable macrostates obtained before, and it represents a TS with large probabilities to jump to the neighboring M-states (Figs. 1a-b), and most of the flux going through the TS.

## III. APPLICATIONS

We apply our clustering algorithm to the analysis of MD trajectories obtained for two different systems (i) the relatively small helix-forming peptide $Ala_5$, and (ii) a larger system, the epidermal growth factor receptor (EGFR) protein (32, 33).

### A. Conformational Dynamics of $Ala_5$

We first use our algorithm to study the dynamics of alanine pentapeptide ($Ala_5$) - a commonly used test system for evaluating the intrinsic conformational kinetics – using atomistic MD trajectories (32, 34). The $Ala_5$ system (Fig. 2) has the advantage of being sufficiently small for generating converged MD sampling, even with an explicit representation of water, using relatively modest computational resources. At the same time, it is the smallest peptide that can form a full helical "*i,i*+4" loop between the amide carbonyl of its first residue and the amide hydrogen of its last residue allowing the study of secondary structure formation in both theoretical (32, 35-38) and experimental (39, 40) studies. We analyzed MD trajectories of $Ala_5$ with TIP3P (41) water molecules initialized from 4 initial conditions, of 250 ns length each, and performed at two different temperatures $T_1 = 300$ K, $T_2 = 350$ K, for a total simulation time of 2 μs. The simulations were performed using GROMACS (42) with the Amber-GSS (43) force field, as described in detail in Ref. (32).

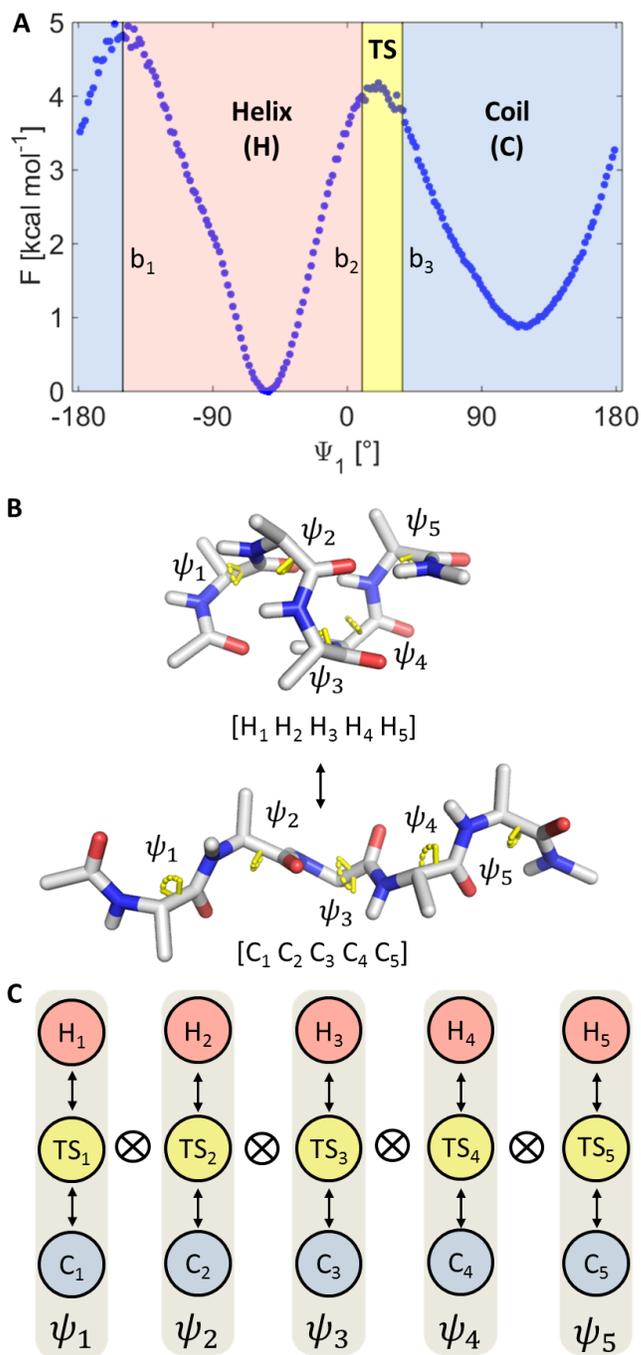

**Figure 2. A** Free energy for the backbone angle $\Psi_1$ of Ala$_5$ calculated for $N_\mu$=200 μ-states at 350 K and lag time τ=1 ps. Resulting M-states (vertical boundary lines: b1, b2, b3) are shown for 3-state clustering with periodic boundaries ((b1, b2) being identical for 2-state clustering). **B** All-helical, 00000, and all-coil, 22222, conformations of Ala$_5$. The five $\Psi$ backbone dihedral angles (yellow) are used as RCs. **C** For *N*-D clustering (*N*=5), μ-state clustering identifies *M*=3 M-states along each RC (H - red, C – blue, and TS – yellow). Here, the final coarse-graining in Ω-states is performed for all the RCs M-states (though it can also be done sequentially, see text).

To cluster the MD trajectory of the Ala$_5$ peptide, the five Ramachandran $\Psi$ angles have been chosen as RCs (Fig. S3), as the free energy barriers along the $\Phi$ angles are much smaller and thus, they contribute less to the slowest relaxation modes (32). In Fig. 2A the clustering result for the 1D profile for $\Psi_1$ is demonstrated, where cluster boundaries are shown along the free energy profile calculated from $N_\mu = 200$ Markov μ-states at 350 K and lag time $\tau = 1$ ps. For all 5 $\Psi$ angles, the two-state assignment successfully identified the two meta-stable states corresponding to the two free energy basins and the slowest relaxation time is in good accordance with the full 200-state model. As demonstrated in Fig. 2A, 3-state clustering for meaningful lag times up to the magnitude of the relaxation time ($\sim\tau = 500$ ps) leads to the detection of the small TS state, which is characterized by a small equilibrium population and survival probability, and similar transition probabilities to either of the H or C states. For example, 3-state clustering of the backbone angle $\Psi_1$ at 350 K (Fig. 2A) with $N_\mu=100$ μ-states leads to a slowest relaxation time of $t_2 = 212.45$ ps, and the equilibrium probabilities are, $\pi_0(H) = 0.662$, $\pi_1(TS) = 0.001$, and $\pi_2(C) = 0.337$. The macrostate boundaries are listed in Table S1, and the transition probabilities are $T_{22} = 0.0984$, $T_{20} = 0.4756$, $T_{21} = 0.4260$.

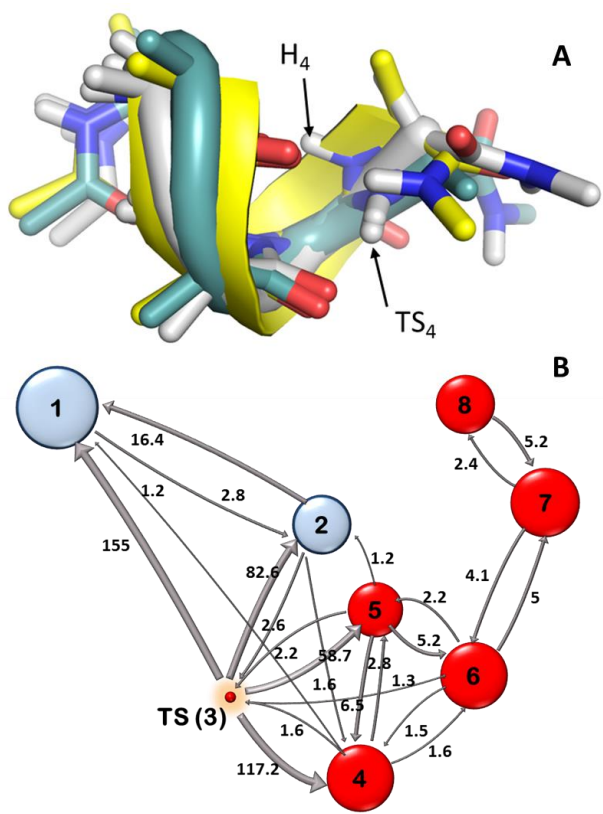

**Figure 3. A** Transitions near the TS are dominated by conformational dynamics at the 4$^{th}$ alanine residue of Ala$_5$. Representative conformations are shown for helical [00000] (yellow cartoon), and TS states [00020] (blue cartoon), and [00021] (grey cartoon). **B** Kinetic network for the optimal coarse grained Ω-states for the Ala$_5$ (350 K, Tables

S1-S3, Fig. S4). Two macrostates ($\Omega_1$ and $\Omega_2$, blue) form the folded ensemble, and five $\Omega$ states (4-8, red) form the unfolded ensemble. The numbers near each arrow are the corresponding transition rates [ns$^{-1}$].

For convenience, we have labelled the $S_i$ M-states with 0 if they are in a helical (H) region of $\Psi$, 1 if they are in a coil (C) region of $\Psi$, and 2 if they are in a TS region of $\Psi$. Thus, the all-helical macrostate of Ala$_5$ is denoted as [00000], while, for example, [02221] will denote an M-state with the N-terminal residue helical, the C-terminal residue coil and the middle residues in their respective TS regions of their $\Psi$ angles.

The 3-state clustering of all 5 $\Psi$ RCs (with boundaries given in Table S1) at 350 K leads to the identification of only 8 $\Omega$ macrostates at lag times $\tau = 10$ ps as depicted in Fig. 3B and presented in Table S2 (including the corresponding $\Omega$-state populations and the equilibrium probabilities of their most populated M-states). The macrostates $\Omega_1$ and $\Omega_2$ have the largest population and form the "folded" ensemble. In this case, $\Omega_1$ consists of only one all-helical M-state $\Sigma_0 = [00000]$ that has an equilibrium population of ~40.84% (Table S2). The 2$^{nd}$ most populated state, $\Omega_2$, consists of several macrostates that add up to 6.91% of the total equilibrium population, of which 00001 is predominant (~6.66%). Thus, unfolding at the last C-terminal residue of Ala$_5$ is not sufficient to perturb the overall stability of helical conformations, and it maintains Ala$_5$ in its folded ensemble.

On the other hand, the unfolded ensemble appears to consist of 5 $\Omega$ states ($\Omega_4$ to $\Omega_8$) connected with noticeably slower rates than those within the folded states, and accounting for more than 52% of the total equilibrium population. The "unfolded" ensemble is separated from the helix-rich "folded" ensemble by a TS, $\Omega_3$, that consists of 10 M-states with a total population of only ~0.19%, of which the most populated M-states are [2000*], [0200*] and [0002*], where the "*" symbol means that at that position either 0 (H) or 1 (C) conformations are observed. Thus, the range of transitions near the TS is largely dominated by conformational dynamics at the 1$^{st}$, 2$^{nd}$, and 4$^{th}$ residue of Ala$_5$. Interestingly, at 300 K the 4$^{th}$ residue is the only residue that significantly modulates the TS dynamics (Table S3).

As illustrated in Fig. 3B, the "unfolded ensemble" states $\Omega_4$ to $\Omega_8$ presents at 350 K (i) a specific connectivity that depends on how many residues can change cooperatively their state, and (ii) transition rates that are lower than in the folded ensemble.

These observations, enabled by our method agree well with previous experimental and computational studies of Ala$_5$ (32), while offering a more detailed, automatic analysis, also identifying TS states.

### B. Dynamics of the EGF Receptor

We also apply our algorithm to analyze the dynamics of the epidermal growth factor receptor (EGFR) transmembrane helices. The MD data was obtained using the Anton supercomputer (1) and was first presented in Ref. (33) in the Shaw group. These simulations include the extracellular module, the transmembrane segment, the juxtamembrane segment, and the intracellular kinase domain separately. Here, we analyze in detail the 100.2 μs trajectory of the N-terminal transmembrane dimer, including the positions of all protein atoms saved with a time step of $\Delta t = 1$ ns at 310 K.

To identify sensible reaction coordinates from the data containing the positions of all 1042 atoms of the protein, time-lagged independent component analysis (tICA) has been used to generate RCs (27, 28). Clustering of the one-dimensional trajectory of the first tICA component (Table S4) leads first to the identification of two metastable M-states (Fig. 4A, red and blue regions), finding subsequently a first TS-like M-state (yellow region between boundaries b1 and b2 in Fig. 4A), followed by another meta-stable state (S3 macrostate in Fig. 4A). The cluster boundaries for 2- to 5-state clustering, together with the slowest relaxation time and the equilibrium distributions of the respective states are summarized in Table S4. The small equilibrium probabilities indicate that the low-populated state is indeed TS-like, which is also confirmed by its relative transition probabilities. A representation of the TS conformation at chain A is illustrated in Fig. 4B, colored according to the mean absolute correlation between each atom coordinate and the first tICA component.

We used the first 4 tICA components to obtain the corresponding global network of transitions between the optimal coarse-grained $\Omega$-states of this EGFR dimer region (Figs. S5-6). The 5$^{th}$ component had a much faster relaxation time (Fig. S7), and thus was not included. The overall network is illustrated in Fig. 4D and presented in more detail in Table S5. Two coarse grained states ($\Omega_1$ and $\Omega_2$) form one conformational basin separated by the TS-like state $\Omega_3$ (orange) from a second basin formed by $\Omega_4$ and $\Omega_5$.

Interestingly, as shown in Table S5, the 2$^{nd}$ tICA component seems to capture best the conformational dynamics between the TS state and neighboring conformations corresponding to subtle conformational changes at the N terminus of chain B (Fig. 4B). The calculated free energy landscapes (Figs. 4A and S6) feature relatively small barriers, which would make any state analysis more challenging. Nevertheless, our algorithm succeeds in identifying the essential TS conformation, $\Omega_3$, that separates the much broader $\Omega_1$ and $\Omega_2$ basins from the ones formed by $\Omega_4$ and $\Omega_5$. Interestingly, to a large extent the $\Omega_4$ basin is quite similar to the $\Omega_3$ TS-like conformation, a result that may change when longer trajectories are available. Our results thus identify key starting points at the TS configurations for additional simulations that are aimed at efficient sampling of the dynamics.

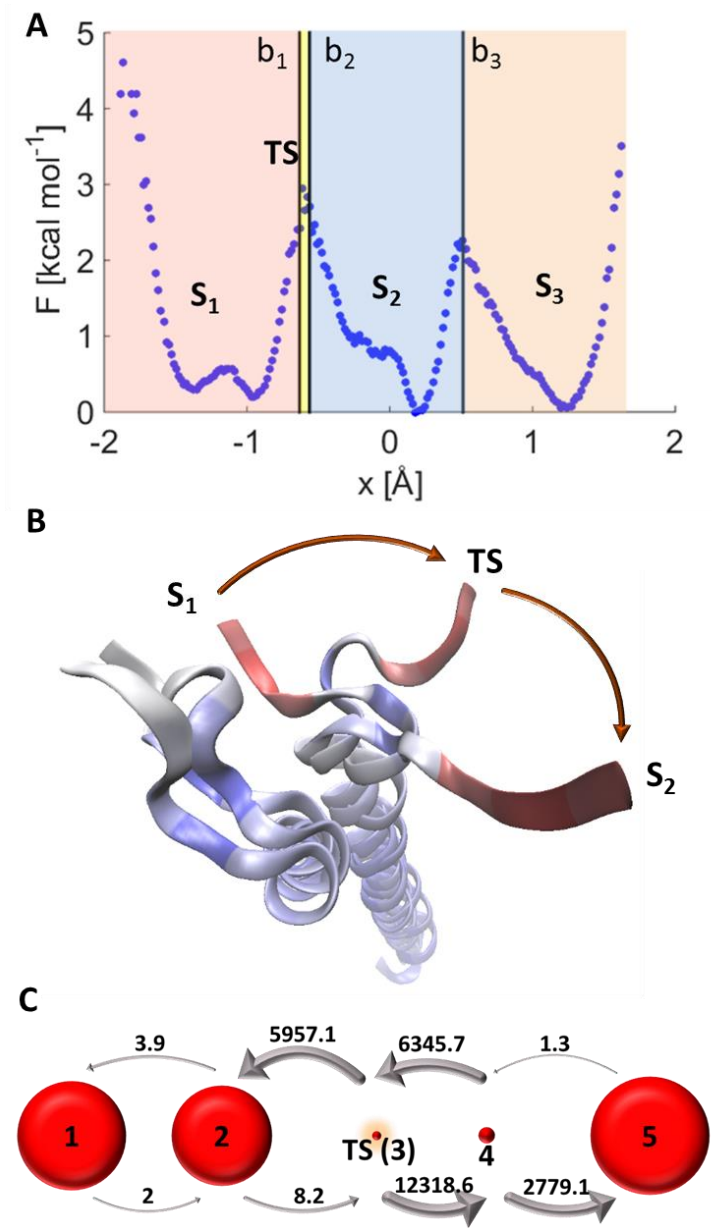

**Figure 4. A** Free energy along the first tICA component of the EGFR transmembrane dimer calculated using an initial 200-state MSM. Results are shown for 4-state (boundaries b1, b2 and b3) clustering for a lag time of $\tau = 1$ ns. Likewise, a 3-state clustering already identifies b1 and b2 boundaries. The TS region (yellow) separates the $S_1$ macrostate from the others. **B** Configurations of the transmembrane module of the EGF receptor that are representative for the $S_1$, TS and $S_2$ regions of the first tICA component at the N-terminal end of chain A. The structures are colored according to their mean absolute correlation with the first tICA component. **C** Global kinetic

network obtained using 4 tICA components at lagtime = 1 ns, kinetic rates are shown in 1/10 μs$^{-1}$. Two coarse grained states ($\Omega_1$ and $\Omega_2$, blue) form one conformational basin separated by the TS-like state $\Omega_3$ (orange) from a second basin formed by $\Omega_4$ and $\Omega_5$ (red).

## IV. CONCLUSIONS

We present a new approach to identify automatically sensible clustering of metastable and transition states along the available reaction coordinates describing the molecular process. An analytical model for 3-state clustering shows two families of solutions to the optimization problem, one of which leads to the identification of metastable states. The second solution yields two metastable and one short-lived state, the latter being identified as a transition state, as the application of the algorithm to analytical Markov models based on arbitrary free energy functions revealed.

The algorithm is presented and studied by using both analytical and actual examples from MD simulations of the helix-forming peptide Ala$_5$, and of a larger system, the epidermal growth factor receptor (EGFR) protein transmembrane region. In all cases, our method identifies automatically the corresponding transition states and metastable conformations in an optimal way, with minimal input, by capturing accurately the intrinsic slowest relaxation times. We find that this new approach provides a general and easy to implement analysis method that provides unique insight into the molecular mechanism and the rare but crucial rate limiting conformational pathways occurring in complex dynamical systems such as molecular trajectories.

In contrast to most other available clustering methods and their applications (44-46) that capture metastable states or high energy intermediates, our approach provides an automatic identification of all important metastable states as well as unstable transition states to describe the slowest process in the system. This type of identification of transition states and metastable states allows us to determine the molecular mechanisms with the key conformational ensembles and could lead to whole new approaches to more efficiently simulate and analyze molecular processes, central to a broad variety of biomolecular research and drug design problems. Our approach is fully general and does not rely on any specific molecular properties in the analysis of time-dependent trajectories, therefore it may also be applicable to time series of complex systems beyond molecular conformations to identify rate limiting events.

## ACKNOWLEDGEMENTS

We thank Drs. Attila Szabo (NIH) and Alessia Annibale (KCL) for numerous helpful discussions, and Albert Pan (D. E. Shaw Research) for his kind assistance with the EGFR simulation data.